\documentclass[11pt,preprint]{aastex}
\slugcomment{Accepted for the publication in {\it The Astrophysical Journal Letters}}
\def\la{\mathrel{\hbox{\rlap{\hbox{\lower4pt\hbox{$\sim$}}}\hbox{$<$}}}}
\def\ga{\mathrel{\hbox{\rlap{\hbox{\lower4pt\hbox{$\sim$}}}\hbox{$>$}}}}

\shortauthors{Park}
\shorttitle{Kepler's SNR}

\begin{document}
\title{A Super-Solar Metallicity for the Progenitor of Kepler's Supernova}

\author{Sangwook Park\altaffilmark{1}, Carles Badenes\altaffilmark{2},
Koji Mori\altaffilmark{3}, Ryohei Kaida\altaffilmark{3}, Eduardo 
Bravo\altaffilmark{4}, Andrew Schenck\altaffilmark{1}, Kristoffer A.
Eriksen\altaffilmark{5}, John P. Hughes\altaffilmark{6}, Patrick O. 
Slane\altaffilmark{7}, David N. Burrows\altaffilmark{8}, and Jae-Joon 
Lee\altaffilmark{9}} 

\altaffiltext{1}{Box 19059, Department of Physics, University of Texas at Arlington,
Arlington, TX 76019; s.park@uta.edu}
\altaffiltext{2}{Department of Physics and Astronomy and Pittsburgh Particle Physics, 
Astrophysics, and Cosmology Center (PITT-PACC), University of Pittsburgh, 3941 O’Hara 
Street, Pittsburgh, PA 15260, USA; badenes@pitt.edu}
\altaffiltext{3}{Department of Applied Physics, University of Miyazaki, 1-1 Gakuen 
Kibana-dai Nishi, Miyazaki, 889-2192, Japan}
\altaffiltext{4}{Department F\'isica i Enginyeria Nuclear, Univ. Polit\'ecnica de 
Catalunya, Carrer Pere Serra 1-15, 08173, Sant Cugat del Vall\'es, Spain}
\altaffiltext{5}{XTD-6, Los Alamos National Laboratory, P.O. Box 1663, Los Alamos, 
NM 87545}
\altaffiltext{6}{Department of Physics and Astronomy, Rutgers University,
136 Frelinghuysen Road, Piscataway, NJ 08854-8019}
\altaffiltext{7}{Harvard-Smithsonian Center for Astrophysics, 60 Garden Street,
Cambridge, MA 02138}
\altaffiltext{8}{Department of Astronomy and Astrophysics, Pennsylvania State
University, 525 Davey Laboratory, University Park, PA 16802}
\altaffiltext{9}{Korea Astronomy and Space Science Institute, Daejeon, 305-348, Korea}


\begin{abstract}

We have performed deep X-ray observations of the remnant of Kepler's supernova 
(SN 1604) as a {\it Key Project} of the {\it Suzaku Observatory}. Our main goal
is to detect secondary Fe-peak elements in the SN ejecta to gain insights into 
the Type Ia supernova explosion mechanism and the nature of the progenitor. 
Here we report our initial results. We made a conclusive detection of X-ray 
emission lines from highly ionized Mn, Cr, and Ni as well as Fe. The observed 
Mn-to-Cr line flux ratio is $\sim$0.60, $\sim$30\% larger than that measured 
in Tycho's remnant. We estimate a Mn-to-Cr mass ratio of $\sim$0.77, which is 
strongly suggestive of a large neutron excess in the progenitor star before 
the onset of the thermonuclear runaway.  The observed Ni-to-Fe line flux 
ratio ($\sim$0.03) corresponds to a mass ratio of $\sim$0.06, which is 
generally consistent with the products of explosive Si-burning regime in 
Type Ia explosion models, and rules out contamination from the products of 
neutron-rich nuclear statistical equilibrium in the shocked ejecta. Together 
with the previously suggested luminous nature of the explosion, these mass 
ratios provide strong evidence for a super-solar metallicity in the SN 
progenitor ($\sim$3$Z_{\odot}$). Kepler's supernova was likely the 
thermonuclear explosion of a white dwarf formed in the recent past that 
must have exploded through a relatively prompt channel.

\end{abstract}


\section {\label {sec:intro} INTRODUCTION}

The progenitors of Type Ia supernovae (SNe) are believed to be C/O white dwarfs 
(WDs) in close binary systems. The WD becomes unstable when it approaches the 
Chandrasekhar limit ($M_{Ch}$ = 1.44 $M_{\odot}$) by either accreting a sufficient 
amount of material from a non-degenerate companion ({\it single-degenerate} 
channel) or merging with another WD ({\it double-degenerate} channel). This mass 
increase triggers a thermonuclear runaway and an explosion ensues by the fusion 
of carbon and oxygen. While this is a generally accepted picture, the nature of 
the progenitor and its binary companion and the details of the explosion process 
remain elusive. The progenitor's nature (e.g., age, mass, and metallicity), its 
environment (e.g., ambient structure and the companion's nature), and the details 
of the explosion mechanism (whether the burning front is a deflagration, a delayed 
detonations, etc.) may significantly vary among SNe Ia. Understanding the diversity 
of SN Ia is essential to reduce systematic uncertainties in their cosmological 
applications. It has been established that SN Ia progenitors have a wide range 
of delay-times, from a few hundred Myr to several Gyr, and that ``prompt'' SNe 
Ia produce more $^{56}$Ni and are brighter than the ``delayed'' population 
\citep[e.g.,][]{maoz12}. If SN Ia exploded with delay times of only a few 
hundred Myr, their progenitor's masses must have been significantly larger 
($\sim$3.5--8 $M_\odot$) than the Sun in their main sequence stage \citep{aubo08}. 

The progenitor's metallicity ($Z$) is a key parameter related to the age 
population of SNe Ia. During the main sequence evolution of an intermediate 
mass star, the CNO-cycle converts C, N, and O into $^{14}$N, which 
subsequently produces $^{22}$Ne through the intermediate ${\beta}^+$ decay 
of $^{18}$F in the star's hydrostatic He-fusion phase. This results in 
neutron excess ($\eta$ = 1--2$Y_e$ where $Y_e$ is the proton to nucleon number 
ratio) in the progenitor. Except for the deepest core of $M$ $<$ 0.2$M_{\odot}$ 
where $\eta$ is dominated by electron-capture in neutron-rich nuclear statistical 
equilibrium (n-rich NSE) \citep{brac00}, $\eta$ is proportional to $Z$ 
\citep{timm03}. During the explosion, the excess amount of neutrons is stored 
in trace elements with unequal number of protons and neutrons.  $^{55}$Mn 
is the most abundant among such elements. $^{52}$Cr is produced by the same 
explosive Si-burning as Mn, but its synthesis is unrelated to $\eta$. Thus, 
the Mn-to-Cr mass ratio is an excellent tracer of the progenitor's $Z$ 
\citep{bade08,brav13}. The progenitor's $Z$, together with the details of 
the explosion physics, can also influence the synthesis of stable Ni 
\citep[mostly $^{58}$Ni, e.g.,][]{iwam99,timm03}. Observational constraints 
on the content of these Fe-group ejecta elements in SNe Ia would thus provide 
insights on the detailed nature of the progenitor system and the explosion 
mechanism. However, the relatively long decay-times of radioactive Fe-peak 
elements \citep[$\sim$yr, e.g.,][]{kuch94} make it difficult to detect these 
elements directly in SNe Ia. Instead young supernova remnants (SNRs) provide 
a useful opportunity.   

Kepler's SNR (``Kepler'' hereafter) is the remnant of SN 1604. The historical 
SN light curve suggested a Type Ia explosion \citep{baad43}, although this was 
not a unique classification \citep{dogg85}. Kepler shows strong atomic X-ray 
emission lines from highly-ionized ions of Si, S, Ca, and Fe 
\citep[e.g.,][]{deco94,kinu99,cass04}. These overall X-ray spectral 
characteristics are similar to those of Tycho's SNR, an unambiguous Type Ia 
\citep{krau08, bade06}. Reynolds et al. (2007) found that the X-ray ejecta are
dominated by Si, S, and Fe, without evidence for O, Ne, and Mg. Based on this 
ejecta abundance structure and the absence of a conspicuous stellar remnant 
\citep[see also][]{blai05}, they ruled out a core-collapse origin for Kepler. 
Kepler appears to be interacting with a modified circumstellar medium 
\citep[CSM,][]{band87}. Overabundant nitrogen was detected in Kepler, 
indicating the presence of dense CSM produced by massive stellar winds 
\citep{denn82,blai91}. The spatial distribution of interacting CSM suggested 
that the strong winds might have originated from an AGB companion 
\citep{chio12,burk13}. These results generally support a single-degenerate 
origin for Kepler. 

Based on observations at several wavebands, a range of distances (from 
$d$ $\sim$ 3 kpc to $d$ $>$ 6 kpc) to Kepler has been suggested 
\citep[e.g.,][]{reyn99,sank05,ahar08,vink08}. Recently, Patnaude et al. 
(2012) showed that hydrodynamic and X-ray spectral simulations are consistent 
with the {\it Chandra} spectrum of Kepler for $d$ $\ga$ 7 kpc if the shock is 
expanding into massive stellar winds, while the distance may be closer ($d$ 
$\sim$ 4 kpc) for the case of a uniform medium. The Type Ia explosion must 
have been bright and $^{56}$Ni-rich to explain the strong Fe line emission 
observed by {\it Chandra} \citep{patn12}.
 
With its Type Ia origin, young age (409 yr), and the ejecta-dominated spectrum 
of X-ray emission, Kepler provides an excellent opportunity to study the details
of Type Ia explosion mechanisms and the progenitor's nature. In this {\it Letter}, 
we report the initial results from our {\it Suzaku Key Project} of Kepler. As
we presented in our preliminary report \citep{park12}, we detect X-ray emission 
from highly-ionized Mn, Cr, Ni, and Fe ions. Line flux ratios of these Fe-group 
elements strongly suggest a super-solar $Z$ for the progenitor. We describe our 
observations in Section~\ref{sec:obs}. Our data analysis and results are presented 
in Section~\ref{sec:result}. A discussion and conclusions are presented in 
Sections~\ref{sec:disc}.  

\section{\label{sec:obs} OBSERVATIONS}

We performed our {\it Suzaku} observations of Kepler as part of a {\it Key Project} 
in 2009 August--2011 March during Cycles 4--5 (Table~\ref{tbl:tab1})\footnote{We 
also used our earlier {\it Suzaku} data (ObsID 502078010), performed on 2008 
February.}. The X-ray Imaging Spectrometer (XIS) was chosen to be the primary 
detector. The FI-CCDs (XIS0 and 3) have a better energy resolution (FWHM $\sim$ 
180 eV at $E$ $\sim$ 6 keV) than the BI-CCD (XIS1, FWHM $\sim$ 250 eV at $E$ $\sim$ 
6 keV). The non-X-ray background flux is significantly lower (by a factor of 
$\sim$2--3 at $E$ = 5--8 keV) in the FI-CCDs than the BI-CCD. When the two 
available FI-CCDs (XIS0 and 3) are combined, the collecting area is larger by a 
factor of 2 than that of XIS1.  Thanks to the good energy resolution and low 
background, the FI-CCDs efficiently detect and measure faint line emission fluxes 
from Fe-group ejecta elements at $E$ $\sim$ 5--8 keV, which is our primary goal. 
Thus, we used the XIS0 and 3 data in this work. Some time intervals in observations 
taken in  March 2011 were excluded from the analysis because of temporary on-board 
software issues with the XIS0 (for $\sim$25 ks in ObsID 505092050). Otherwise, we 
followed the standard data reduction process including charge injection, and 
combined all 3$\times$3 and 5$\times$5 mode spectra for each XIS0 and XIS3. After 
the data reduction, the total effective exposures are $\sim$665 and $\sim$691 ks 
for XIS0 and XIS3, respectively.

Since Kepler is a bright extended source ($\sim$2$'$ in radius), scattered photons 
from the SNR dominate nearly the entire CCD due to the large point spread function 
of the telescope, which makes it difficult to adequately estimate the background 
spectrum from surrounding regions. The characterization of the background spectrum 
is critical to accurately measure fluxes from faint line features. To mitigate the 
bright scattered light from the SNR, we have performed separate background observations. 
We chose four nearby source-free regions ($\sim$1.5$^{\circ}$ from Kepler), one each 
of which is to north, south, east, and west of Kepler (Table~\ref{tbl:tab1}). We 
processed these background observations in the same way as we did for the source 
observations. After the data reduction, the total effective exposure for the 
background observations is $\sim$273 ks each for XIS0 and XIS3. We note that 
recently Yang et al. (2013) analyzed our {\it Suzaku} data of Kepler using 
background estimates from source-free regions on the same XIS chip instead of 
using separate background pointings. Their measurements of lower fluxes (by 
$\sim$40\%) for Mn K$\alpha$ and Cr K$\alpha$ lines than ours (see 
Section~\ref{sec:result}) are likely caused by the scattered X-ray contamination 
in their background estimates. They suggested a carbon-deflagration for Kepler's
explosion based on a low value of the Cr-to-Fe K$\alpha$ line equivalent width
(EW) ratio ($\sim$0.008). Our larger Cr line flux indicates the Cr-to-Fe K$\alpha$ 
line EW ratio of $\sim$0.016 which would rather favor delayed detonations for 
Kepler as are self-consistently suggested by our Ni-to-Fe mass ratio measurement
(Section~\ref{sec:disc}).

\section{\label{sec:result} ANALYSIS \& RESULTS}

We extracted the source spectrum from the entire SNR (a circular region of 4$'$ 
in radius). We extracted the background spectrum from the same detector region 
as the source in each background pointing, and combined them. In the 5--8 keV band, 
we obtained $\sim$130000 counts from the source for each of XIS0 and XIS3. The 
background contribution is $\sim$7\%. For the spectral model fits, we re-binned 
the source spectrum to contain a minimum of 50 counts per energy channel. Our 
main goal is to measure line fluxes from Cr, Mn, Fe, and Ni using the integrated 
X-ray spectrum of Kepler. Thus, we fit the background-subtracted spectrum with a 
phenomenological model in the 5--8 keV band. We fit the observed spectrum with
an absorbed power law (PL) model with 5 Gaussian components. The PL component is to
characterize the underlying continuum. Each Gaussian is to fit the emission feature 
from Cr K$\alpha$ ($E$ $\sim$ 5.5 keV), Mn K$\alpha$ ($E$ $\sim$ 5.9 keV), Fe 
K$\alpha$ ($E$ $\sim$ 6.5 keV), Fe K$\beta$ ($E$ $\sim$ 7.1 keV), and Ni K$\alpha$ 
($E$ $\sim$ 7.5 keV) lines, respectively. The same Gaussian line width was assumed 
for all five lines, and was varied to give the best fit. The foreground column is 
fixed at $N_{\rm H}$ = 5.2 $\times$ 10$^{21}$ cm$^{-2}$ \citep{reyn07}. The photon 
index ($\Gamma$) and normalization parameters for the PL and Gaussian components 
are varied freely. We fit the XIS0 and XIS3 spectra simultaneously, tying all 
parameters between the two spectra. 

\begin{figure}[]
\figurenum{1}
\centerline{\includegraphics[angle=0,width=\textwidth]{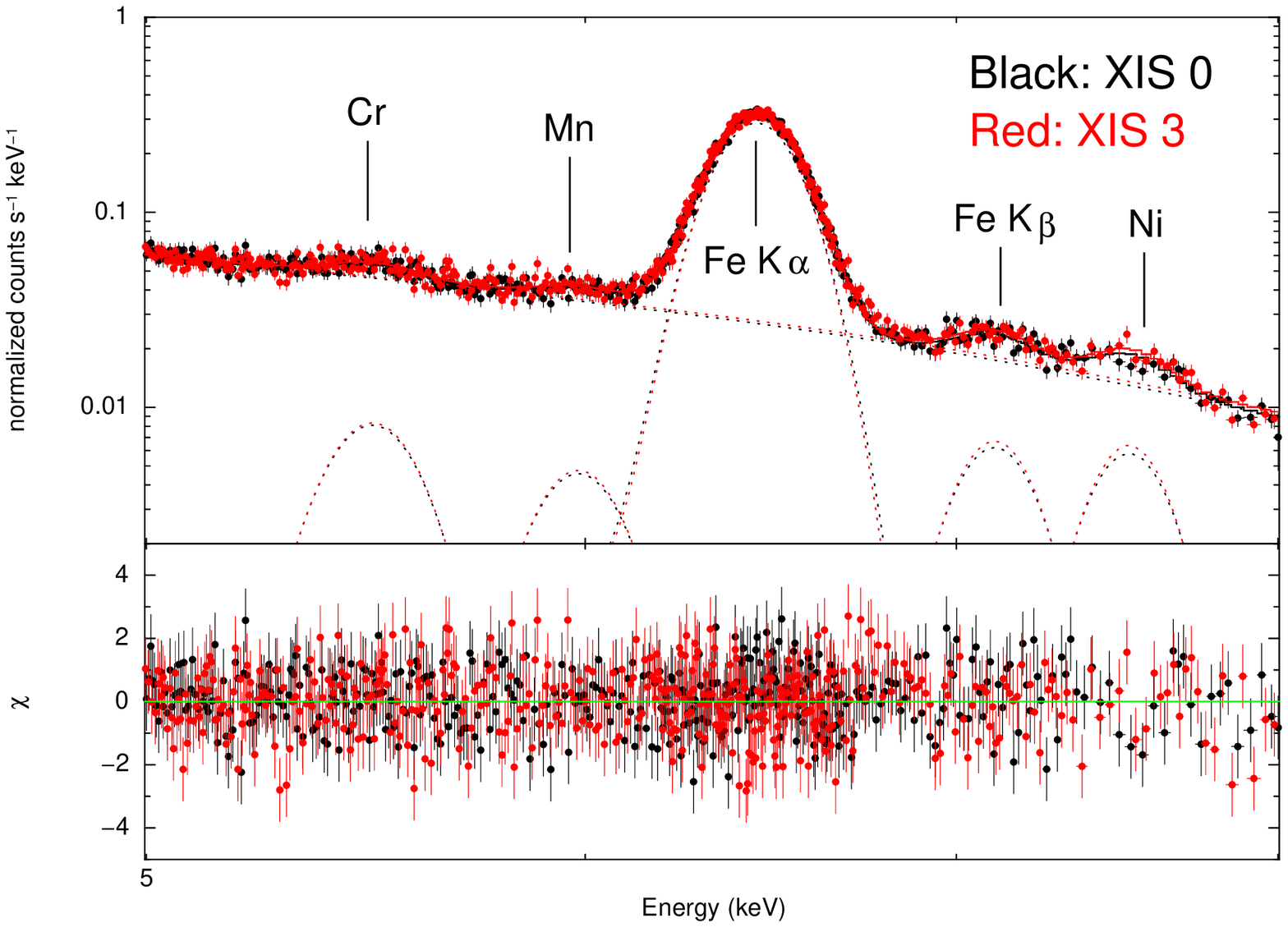}}
\figcaption[]{{\it Suzaku} XIS spectrum of Kepler's SNR. The best-fit model 
(PL + 5 Gaussians) are overlaid. 
\label{fig:fig1}}
\end{figure}

The observed spectrum and the best-fit model ($\Gamma$ = 2.60, $\chi^2$/$\nu$ 
= 1469.8/1513) are shown in Figure~\ref{fig:fig1}. In the 5--8 keV band, X-ray 
emission is dominated by the strong Fe K$\alpha$ line at $E$ $\sim$ 6.45 keV. 
X-ray emission lines from He-like ions of Mn and Cr are clearly detected (at 
$\sim$9$\sigma$ and $\sim$14$\sigma$ level for Mn and Cr, respectively) 
(Table~\ref{tbl:tab2} \& Figure~\ref{fig:fig1}). Fe K$\beta$ ($E$ $\sim$ 
7.1 keV) and Ni K$\alpha$ ($E$ $\sim$ 7.5 keV) lines are also conclusively 
detected. To verify that the Ni line was not due to the instrumental Ni 
fluorescent line, which varies with the incidental cosmic-ray flux, we 
repeated our Ni line flux measurement using an off-source background region 
from the on-target data. We found that the Ni line was clearly present with 
this background subtraction. We also compared the instrumental Au Ly$\alpha$ 
line flux ($E$ = 9.671 keV) between the source and background observations. 
These tests indicated only a $\sim$10--30\% difference in the instrumental 
line fluxes between the source and background pointings. Thus, we conclude 
that the detected Ni line originates from Kepler. The small difference in 
the instrumental background does not affect our discussion in 
Section~\ref{sec:disc}. Alternatively, we fit the underlying continuum 
with a thermal bremsstrahlung model ($kT$ = 5.11 keV, $\chi^2$/$\nu$ 
= 1465.2/1513).  The estimated fluxes for Fe and Ni lines are identical 
to those from the PL model fit. Line fluxes from Mn and Cr are estimated 
to be slightly smaller (by $\sim$10\%), but the flux ratio is consistent 
(within 1$\sigma$ uncertainties) with that estimated by the PL model fit. 

The archival {\it Chandra} data of Kepler show thin non-thermal filaments along 
the outermost boundary, which are identified as synchrotron radiation from the 
shock-accelerated relativistic electrons \citep{reyn07}. To test the effect from 
synchrotron emission in the measured line fluxes, we repeated our spectral model 
fit with a PL + thermal bremsstrahlung + 5 Gaussians. Based on PL model fits of 
several non-thermal filaments using the archival {\it Chandra} data of Kepler, 
we estimated $\Gamma$ $\sim$ 2.4--2.6 for these features. Thus, we performed the 
model fits with $\Gamma$ fixed at 2.4, 2.5, and 2.6. Using the archival {\it 
Chandra} data of Kepler, we estimated that the thin non-thermal shell (typically 
with a width of $\sim$5$^{\prime\prime}$) all around the SNR contributes 
$\sim$10\% of photon counts in the 5--8 keV band. Thus, we assumed a 10\% 
fractional contribution from the PL component in the 5--8 keV band total flux. 
These model fits are statistically good ($\chi^2$/$\nu$ $\sim$ 1.0) with 
negligible changes in the best-fit parameters from those described above. The 
best-fit thermal bremsstrahlung temperature is $kT$ = 5.1 keV, and the estimated 
line fluxes are consistent with the results derived above. We conclude that our 
measurements of individual line fluxes are robust, and hereafter our discussion 
is based on the results listed in Table~\ref{tbl:tab2}. 

\section{\label{sec:disc} DISCUSSION \& CONCLUSIONS}

Our deep Suzaku XIS data allow us to accurately measure the Fe K$\alpha$ line
center (Table 2), which constrains the ionization timescale to $n_et$
$\approx$ 2 ($\pm$0.03) $\times$ 10$^{10}$ cm$^{-3}$ s for temperatures in 
the range $kT$ = 3--8 keV for the Fe-zone ejecta gas. The estimated Mn-to-Cr 
flux ratio is $f_{\rm Mn}$/$f_{\rm Cr}$ = 0.60$\pm$0.16 (2$\sigma$ uncertainties, 
hereafter). This flux ratio is $\sim$30\% larger than that inferred for 
Tycho \citep{bade08,tama09}. The Mn-to-Cr mass ratio can be calculated using 
their line flux ratio: $M_{\rm Mn}$/$M_{\rm Cr}$ = 1.057 ($f_{\rm Mn}$/$f_{\rm 
Cr}$)/(${\varepsilon}_{\rm Mn}$/${\varepsilon}_{\rm Cr}$), where 
${\varepsilon}_{\rm Mn}$/${\varepsilon}_{\rm Cr}$ is the ratio of specific 
emissivities per ion, and 1.057 is the atomic mass ratio between Mn and Cr 
\citep{bade08}. We calculate ${\varepsilon}_{\rm Mn}$/${\varepsilon}_{\rm 
Cr}$ = 0.82$\pm$0.20 appropriate for the plasma with $kT$ = 3--8 keV and 
$n_et$ = 2 $\times$ 10$^{10}$ cm$^{-3}$ s using the new atomic data (Eriksen
et al. in preparation) generated for this project with Flexible Atomic Code 
\citep{gu08}. The uncertainty of ${\varepsilon}_{\rm Mn}$/${\varepsilon}_{\rm 
Cr}$ depends on the measurements on the gas temperature, ionization timescale, 
and the atomic data. Based on our inferred ranges of $kT$ and $n_et$ as well 
as the uncertainties of the atomic data, we estimate $\sim$20--25\% 
uncertainties on ${\varepsilon}_{\rm Mn}$/${\varepsilon}_{\rm Cr}$. Thus, 
we adopted a conservative limit of a 25\% uncertainty on ${\varepsilon}_{\rm 
Mn}$/${\varepsilon}_{\rm Cr}$. Then, we estimate  $M_{\rm Mn}$/$M_{\rm Cr}$ 
= 0.77$^{+0.53}_{-0.31}$. 

\begin{figure}[]
\figurenum{2}
\centerline{\includegraphics[angle=0,width=\textwidth]{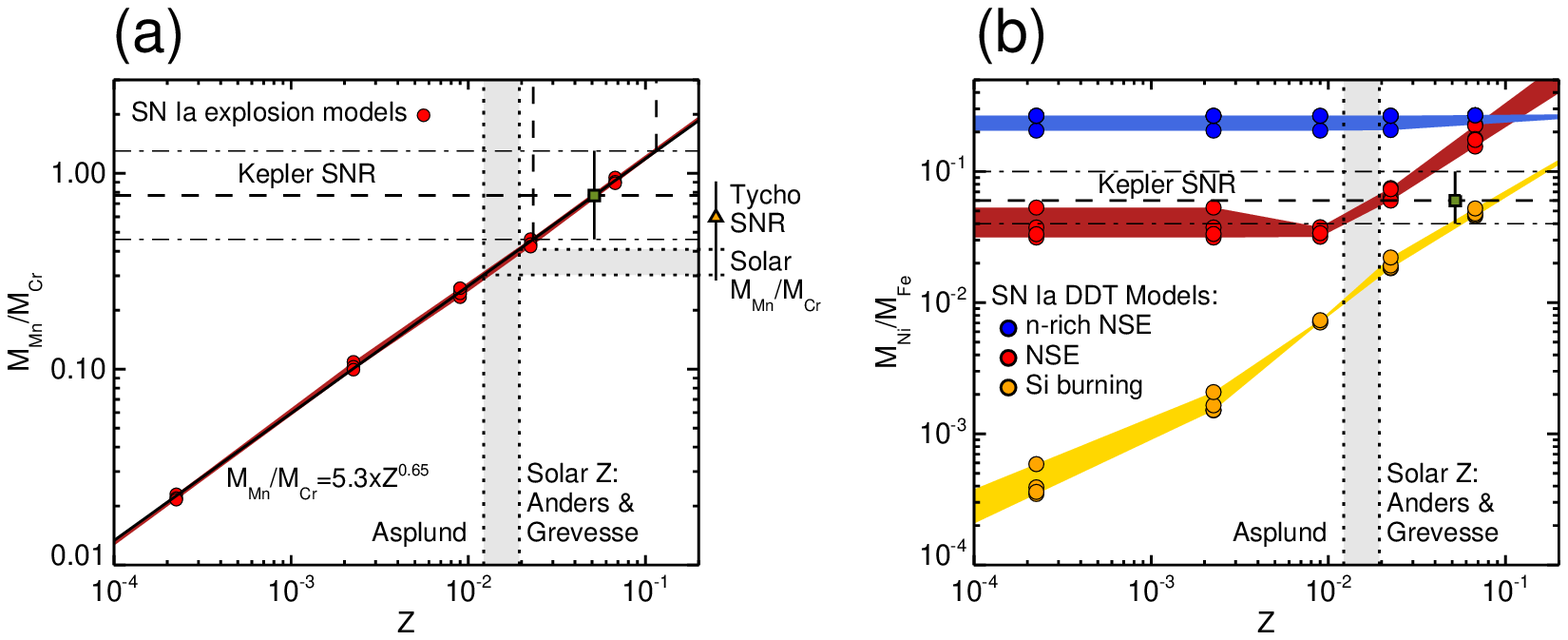}}
\figcaption[]{(a) $M_{\rm Mn}$/$M_{\rm Cr}$ as a function of $Z$ in SN Ia 
explosion models \citep[adapted from][]{bade08}. The models include 1D and 
3D simulations for all major explosion mechanisms \citep[see][]{bade08}. 
The range of solar $Z$ and the corresponding $M_{\rm Mn}$/$M_{\rm Cr}$ are 
shown by the grey area. Our measured $M_{\rm Mn}$/$M_{\rm Cr}$ for Kepler 
is marked. For comparison, $M_{\rm Mn}$/$M_{\rm Cr}$ for Tycho's SNR is also 
shown to the right of the plot. (b)  $M_{\rm Ni}$/$M_{\rm Fe}$ as a function 
of progenitor $Z$ for different burning regimes for delayed detonation SN Ia
explosion models \citep[the same models from][]{bade08}. Our estimate for 
Kepler is shown.
\label{fig:fig2}}
\end{figure}

Assuming that the neutron excess in Kepler's progenitor was not significantly 
affected by the innermost NSE region, we may estimate the progenitor's $Z$ 
using a PL relation betweeen $M_{\rm Mn}$/$M_{\rm Cr}$ and $Z$ \citep{bade08}: 
i.e., unless the non-explosive C-burning, the so-called  C-{\it simmering}
\citep{piro08}, was widespread shortly before the SN explosion, $Z$ shows a 
simple relationship of $M_{\rm Mn}$/$M_{\rm Cr}$ = 5.3$Z^{0.65}$ \citep{bade08}.
While C-{\it simmering} may modify the neutron excess in sub-luminous SNe, 
where the explosive Si-burning region extends deeper into the ejecta 
\citep{bade08}, there is no evidence for a significantly sub-luminous explosion 
for Kepler in the historical light curve. Recent distance estimates of $d$ $\ga$ 
6 kpc \citep{ahar08,patn12} support a normal or luminous SN for Kepler. The {\it 
Chandra} X-ray spectrum of Kepler is consistent with an explosion model that 
produced $\sim$1 $M_{\odot}$ of Fe \citep{patn12}. Thus, Kepler is unlikely the 
remnant of a sub-luminous SN.  Then, we estimate that the progenitor's metallicity 
is $Z/Z_{\odot}$ = 3.6$^{+4.6}_{-2.0}$ for $Z_{\odot}$ = 0.014 \citep{aspl09}, 
and 2.7$^{+3.4}_{-1.5}$ for $Z_{\odot}$ = 0.019 \citep{ande89}. This result 
provides the first observational evidence that the progenitor of a SN Ia was 
a star with a super-solar $Z$ (Figure~\ref{fig:fig2}a). 

We detect the Ni K$\alpha$ line for the first time in Kepler. Since the bulk of 
Fe is created by the radioactive decay of $^{56}$Ni, $M_{\rm Ni}$/$M_{\rm Fe}$ 
$\approx$ $M_{\rm ^{58}Ni}$/$M_{\rm ^{56}Ni}$ = 1.051 ($f_{\rm Ni}$/$f_{\rm 
Fe}$)/(${\varepsilon}_{\rm Ni}$/${\varepsilon}_{\rm Fe}$), where 1.051 
is the atomic mass ratio between $^{58}$Ni and $^{56}$Fe. For the measured
$f_{\rm Ni}$/$f_{\rm Fe}$ = 0.031$\pm$0.005 and ${\varepsilon}_{\rm 
Ni}$/${\varepsilon}_{\rm Fe}$ = 0.51$\pm$0.13 for the K$\alpha$ line in
the same plasma condition discussed above, we estimate $M_{\rm ^{58}Ni}$/$M_{\rm 
^{56}Ni}$ $\sim$ 0.06$^{+0.04}_{-0.02}$. This Ni mass ratio is in good agreement 
(within $\sim$10\%) with delayed detonation models, while being significantly 
different (by a factor of $>$2) from deflagration models \citep{iwam99}. 

Our measured $M_{\rm Ni}$/$M_{\rm Fe}$ indicates that the bulk of the shocked 
Fe-group ejecta was synthesized in the explosive Si-burning regime, especially 
considering the high $Z$ inferred from $M_{\rm Mn}$/$M_{\rm Cr}$ 
(Figure~\ref{fig:fig2}b). Contamination by a small amount of NSE material is 
possible, but a large contribution of products from the n-rich NSE regime deep 
in the core of the exploding star is ruled out, regardless of the $Z$ value. 
This strengthens our confidence on $M_{\rm Mn}$/$M_{\rm Cr}$ as a tracer of 
the neutron excess in the progenitor (and hence $Z$), since it is unlikely that 
material synthesized in the inner n-rich core can contaminate the shocked ejecta 
and increase $M_{\rm Mn}$/$M_{\rm Cr}$ while keeping $M_{\rm Ni}$/$M_{\rm Fe}$ 
at its observed value.

The observed K$\alpha$ line flux ratio $f_{\rm Ni}$/$f_{\rm Fe}$ is higher 
(by $\sim$25\%) in the northern half of Kepler than in the southern half, while 
it is identical (within $\sim$2\%) between the eastern and western halves of 
the SNR. The Ni K$\alpha$-to-Fe K$\beta$ line flux ratio is also 
$\sim$70\% higher in the northern half than in the south, while it is fully 
consistent between the east and west (within $\sim$2\%). The presence of the 
bright bow shock-like X-ray emission feature in the northern shell of Kepler 
suggests a more efficient development of the reverse-shock there than in the 
southern shell. The reverse-shock might have progressed further into the SNR 
in the northern shell (than in the south) to shock more Ni-rich ejecta close 
to the SNR center. Thus, although the spatial variation of the measured Ni-to-Fe 
line flux ratios is statistically marginal ($\sim$1.5--2.5$\sigma$), it would 
not be surprising if such a variation has a real physical  origin. On the other 
hand, we do not find evidence for a similar spatial variation in the Mn-to-Cr 
line flux ratio. The Mn-to-Cr line flux ratio is consistent within 1$\sigma$ 
uncertainties between the north/south and the east/west halves of the SNR. 
This suggests that the correlation between the observed Mn-rich and Ni-rich 
ejecta is probably insignificant even if the Ni-rich ejecta were produced in 
the NSE. Thus, the bulk of the detected neutron excess appears to be associated 
with material synthesized in the explosive Si-burning regime of a relatively 
bright SN Ia, and it should be a tracer of the progenitor's $Z$. 

The implied high $Z$ for the progenitor is consistent with the location of 
Kepler close to the Galactic center, where the interstellar chemical composition 
is expected to further enriched than in the solar neighborhood. Since higher-$Z$ 
stars likely result in a lower-mass core \citep{weis09}, Kepler's progenitor WD 
probably went through an efficient mass growth before exploding as a relatively 
young prompt population SN Ia. A quantitative assessment of the suggested spatial 
variation of the Ni-to-Fe mass ratio and the effect from the core electron-capture 
on the observed neutron excess may require detailed hydrodynamic calculations 
beyond the scope of this {\it Letter}.

\acknowledgments

The authors thank H. Kai and T. Yoshidome for their help with the XIS 
instrumental background investigation. This work has been supported in part 
by the NASA grants NNX09AV42G and NNX10AR47G. The work of K.M. has been partially 
supported by the Grant-in-Aid for Young Scientists (B) of the MEXT (No. 24740167).
P.O.S. acknowledges partial support from NASA Contract NAS8-03060

\clearpage

\begin{deluxetable}{ccccc}
\footnotesize
\tablecaption{{\it Suzaku} Observation Log of Kepler's Supernova Remnant
\label{tbl:tab1}}
\tablewidth{0pt}
\tablehead{\colhead{ObsID} & \colhead{Date} & \colhead{Target} & \colhead{Pointing} & 
\colhead{Exposure} \\
\colhead{} & \colhead{} & \colhead{Name} & \colhead{($\alpha_{2000}$[$^{\circ}$], 
$\delta_{2000}$[$^{\circ}$])} & \colhead{(ks)}} 
\startdata
502078010 & 2008-2-18 & Kepler & 262.6757, -21.4668 & 117.0 \\
504101010 & 2009-9-13 & Kepler\_BG\_GE & 263.5352, -20.2853 & 47.0 \\
504101020 & 2009-8-29 & Kepler\_BG\_GE & 263.5352, -20.2853 & 24.7 \\
504103010 & 2009-10-1 & Kepler\_BG\_GN & 261.9150, -20.2528 & 68.6 \\
504104010 & 2009-10-3 & Kepler\_BG\_GS & 263.1005, -22.9710 & 67.3 \\
504102010 & 2009-10-5 & Kepler\_BG\_GW & 261.8058, -22.7903 & 65.3 \\
505092010 & 2010-9-30 & Kepler SNR & 262.6746, -21.5263 & 17.7 \\
505092020 & 2010-10-6 & Kepler SNR & 262.6739, -21.5293 & 111.2 \\
505092030 & 2011-2-23 & Kepler SNR & 262.6746, -21.4606 & 34.2 \\
505092040 & 2011-2-28 & Kepler SNR & 262.6730, -21.4605 & 146.2 \\
505092050 & 2011-3-8 & Kepler SNR & 262.6732, -21.4607 & 84.6 \\
505092060 & 2011-3-14 & Kepler SNR & 262.6734, -21.4607 & 46.5 \\
505092070 & 2011-3-29 & Kepler SNR & 262.6738, -21.4631 & 133.4 \\
\enddata
\end{deluxetable}

\begin{deluxetable}{cccc}
\footnotesize
\tablecaption{Summary of Detected Atomic Emission Lines
\label{tbl:tab2}}
\tablewidth{0pt}
\tablehead{ \colhead{Line Center Energy} & \colhead{Line Width\tablenotemark{a}} 
& \colhead{Line Flux} & \colhead{Element} \\
\colhead{(keV)} & \colhead{(eV)} & \colhead{(10$^{-6}$ photon 
cm$^{-2}$ s$^{-1}$)}} 
\startdata
5.50$\pm$0.02 & 65.5$\pm$1.2 & 8.31$\pm$1.20 & Cr (K$\alpha$ line) \\
5.99$\pm$0.04 & 65.5$\pm$1.2 & 4.99$\pm$1.16 & Mn (K$\alpha$ line) \\
6.449$\pm$0.0008 & 65.5$\pm$1.2 & 365.91$\pm$2.78 & Fe (K$\alpha$ line) \\
7.12$\pm$0.02 & 65.5$\pm$1.2 & 9.92$\pm$1.24 & Fe (K$\beta$ line) \\
7.53$\pm$0.02 & 65.5$\pm$1.2 & 11.18$\pm$1.88 & Ni (K$\alpha$ line) \\
\enddata
\tablecomments{2$\sigma$ uncertainties are shown.}
\tablenotetext{a}{The line width is tied among all lines.}
\end{deluxetable}

\end{document}